# Baseline glycemia exhibits non-random, history-dependent variation across repeated meals

Arturo Tozzi (corresponding author)
ASL Napoli 1 Centro, Distretto 27, Naples, Italy
Via Comunale del Principe 13/a 80145
tozziarturo@libero.it

ABSTRACT

Glycemic regulation is often described as maintaining glucose levels near a stable baseline. However, continuous glucose monitoring after meals displays intra-individual variability even under controlled conditions, suggesting intrinsic system dynamics beyond sensor noise, measurement error or short-term variability around a fixed set point. Therefore, we estimated pre-meal glucose baselines, tracking their changes across repeated identical meal challenges within individuals. The baseline was defined as the median glucose level in a pre-meal window, while successive displacements were computed between consecutive repetitions. Using a publicly available dataset of normoglycemic subjects, we observed systematic changes in baseline levels across repeated exposures. These displacements exceeded short-term fluctuations within the same pre-meal interval and were robust to alternative baseline definitions. Moreover, the magnitude of each baseline shifted is positively related to the size of the preceding postprandial response. This association persisted under permutation testing, indicating that it cannot be explained by random temporal ordering. Overall, these findings suggest that glycemic dynamics cannot be fully described as independent fluctuations around a fixed baseline. Instead, baseline levels evolve across repeated perturbations through history-dependent adjustments, such that each perturbation influences subsequent system states. Potential applications include refined interpretation of continuous glucose monitoring data and development of models that incorporate temporal dependence in glucose dynamics.

INTRODUCTION

Glycemic regulation is commonly assessed in terms of a homeostatic process in which blood glucose returns to a stable baseline after perturbations such as food intake (Hall et al. 2018; Gregorio et al. 2021; Zhang et al. 2023; Panigrahi and Mohanty 2023; Guo et al. 2025). This view is supported by physiological models based on feedback control involving insulin secretion, hepatic glucose production and peripheral uptake and is widely used in clinical and computational approaches that quantify variability as dispersion around a fixed set point (Xie et al. 2019; Ehrlich et al. 2021; Lewis et al. 2021; Jung et al. 2022; Norton et al. 2022; Tao et al. 2022; Scoditti et al. 2024). However, continuous glucose monitoring has revealed substantial intra-individual variability even under controlled conditions, including different responses to identical meals (Lu et al. 2024; Ferreira et al. 2024; Horgan et al. 2024; Ajjan et al. 2024; Zahalka et al. 2024; Chai et al. 2025). This variability is usually attributed to stochastic factors, measurement noise, or unobserved contextual influences, implicitly assuming that the underlying baseline is stationary and that fluctuations are independent across events. Within this perspective, current descriptions of glycemic dynamics rarely consider the possibility that variability may be structured, history-dependent, and expressed at the level of the baseline itself. However, principles from learning theory such as the law of effect suggest that prior outcomes may shape subsequent responses (Wasserman, Orr, and Li 2026), raising the possibility that biological dynamical systems, including glycemic regulation, may retain traces of previous perturbations rather than fully resetting after each event.

We introduce a quantitative approach to test this hypothesis by focusing on the dynamics of pre-perturbation baselines across repeated identical challenges. We define an effective baseline as a summary statistic of glucose values within a pre-meal temporal window and track its displacement across successive exposures within individuals. This allows variability to be separated into short-scale fluctuations within each interval and larger-scale shifts between consecutive baselines. By comparing the magnitude of baseline shifts with local variability and by relating these shifts to preceding postprandial excursions, our approach evaluates whether baseline changes are consistent with independent noise or instead reflect structured dependence on prior states. Applied to controlled repeated-meal data in normoglycemic individuals, we aim to test whether baseline displacements exceed local variability and depend on the magnitude of preceding excursions, thereby assessing whether glycemic regulation returns to a single invariant level or instead evolves across successive states shaped by recent perturbations.

We will proceed as follows. First, we describe the dataset, inclusion criteria and preprocessing steps. Then, we define the baseline estimation procedure and displacement metrics together with the statistical analyses. Finally, we present the results and assess their implications.

## METHODS

We analyzed glycemic time series from controlled repeated meal challenges to test whether pre-meal glucose baselines remain stable or change across repeated exposures. We used continuous glucose monitoring data together with subject-level information to select strictly normoglycemic individuals and identify repeated identical meals within each subject. For each meal, we estimated a local pre-meal baseline, measured how this baseline changed from one exposure to the next, and quantified post-meal responses using incremental area. We then compared the size of these baseline changes with short-term fluctuations and tested whether they were related to the magnitude of the previous post-meal response. Our goal was to determine whether baseline shifts are simply due to random variability or instead reflect a structured, history-dependent process, in which prior perturbations influence subsequent baseline levels.

**Dataset description.** Our analysis was conducted on the publicly available Stanford Continuous Glucose Monitoring Database, which comprises glucose time series collected under controlled repeated meal-challenge conditions. See also: W et al. 2025. The dataset includes approximately 55 participants spanning a range of metabolic phenotypes. Glucose measurements are recorded at a sampling interval of approximately 5 minutes and expressed in mg/dL, with each observation indexed by subject identifier, food condition, repetition number and time relative to meal ingestion, denoted $t = \text{mins\_since\_start}$, covering the interval $t \in [-25, 170]$ minutes. A companion metadata table provides subject-level variables including glycated hemoglobin (HbA1c, %), fasting glucose (mg/dL) oral glucose tolerance test at 120 minutes (OGTT120, mg/dL), body mass index (kg/m²), age (years) and sex. Subjects were included if they satisfied the criteria HbA1c < 5.7, $G_{\text{fasting}} < 100$ mg/dL and $G_{\text{OGTT120}} < 140$ mg/dL, yielding a subset of 14 normoglycemic individuals (10 female, 4 male) with mean age $54.26 \pm 8.84$ years, mean body mass index $26.50 \pm 3.21$ kg/m², mean HbA1c $5.27 \pm 0.33$%, mean fasting glucose $88.10 \pm 6.80$ mg/dL and mean OGTT120 $103.36 \pm 20.15$ mg/dL. Within this subset, repeated identical meal challenges were available for each subject, allowing construction of 67 repeated meal pairs defined by consecutive repetitions of the same food condition within the same individual. The dataset structure enables direct within-subject comparisons under controlled inputs, with each trajectory represented as $G_{i,f,r}(t)$, where $i$ indexes subjects, $f$ food conditions and $r$ repetition number.

**Preprocessing steps.** Glucose time series were first aligned relative to meal onset by using the provided variable $t = \text{mins\_since\_start}$. Only segments with $t \in [-25, 170]$ minutes were retained to ensure consistent temporal coverage across repetitions. For each subject $i$, food condition $f$ and repetition $r$, the glucose signal is denoted $G_{i,f,r}(t)$. Missing values were excluded without interpolation to avoid introducing artificial smoothness. Data were grouped by subject and food condition and repetitions were sorted in ascending order of $r$. All computations were performed within these grouped structures to preserve within-subject and within-condition consistency. Preprocessing ensures that each repetition corresponds to a comparable temporal window, allowing direct computation of baseline quantities and postprandial responses. Units were preserved in mg/dL throughout all transformations. The resulting dataset consists of a collection of aligned trajectories $\{G_{i,f,r}(t)\}$, each defined on a discrete set of time points, with consistent indexing across subjects and repeated conditions.

**Baseline estimation.** For each trajectory $G_{i,f,r}(t)$, a local pre-perturbation baseline was defined as the median glucose concentration within a fixed temporal window preceding meal onset. Specifically, the baseline is given by
$$G^*_{i,f,r} = \text{median}\{G_{i,f,r}(t): t \in [-25, -5]\}.$$

The use of the median reduces sensitivity to transient fluctuations and measurement noise within the pre-meal interval. In addition to the central estimate, local variability was quantified by the sample standard deviation
$$\sigma_{i,f,r} = \sqrt{\frac{1}{N-1}\sum_t (G_{i,f,r}(t) - \bar{G}_{i,f,r})^2},$$

where the sum is taken over the same pre-meal window and $\bar{G}_{i,f,r}$ denotes the mean in that interval. These quantities provide a characterization of the baseline level and its short-scale variability. Baseline estimates were computed independently for each repetition, allowing comparison across repeated identical conditions. This procedure yields a sequence $\{G^*_{i,f,r}\}$ indexed by repetition number, forming the basis for subsequent displacement analysis.

**Displacement metric.** Successive baseline displacement was defined for each subject and food condition as the difference between consecutive baseline estimates. For repetitions $r$ and $r-1$, the displacement is given by
$$D_{i,f,r} = G^*_{i,f,r} - G^*_{i,f,r-1}.$$

The absolute displacement magnitude is
$$|D_{i,f,r}| = |G^*_{i,f,r} - G^*_{i,f,r-1}|.$$

To assess whether these displacements exceed local variability, a pooled baseline standard deviation was defined as

$$\sigma_{i,f,r}^{\text{pool}} = \sqrt{\frac{\sigma_{i,f,r}^2 + \sigma_{i,f,r-1}^2}{2}}.$$

Comparisons were then made between $|D_{i,f,r}|$ and $\sigma_{i,f,r}^{\text{pool}}$ and $2\sigma_{i,f,r}^{\text{pool}}$ to determine whether displacements exceed one or two standard deviations of local baseline variability. This formulation provides a normalized measure of displacement relative to short-scale fluctuations and allows evaluation of whether observed differences are compatible with local noise.

**Excursion quantification.** Postprandial responses were quantified using the incremental area under the curve (iAUC) above baseline. For each trajectory, the incremental signal is defined as

$$\Delta G_{i,f,r}(t) = \max\left(G_{i,f,r}(t) - G_{i,f,r}^*, 0\right).$$

The iAUC over the interval $t \in [0,120]$ minutes is computed using trapezoidal integration:

$$\text{iAUC}_{i,f,r} = \int_0^{120} \Delta G_{i,f,r}(t)\, dt \approx \sum_k \frac{\Delta G_{k+1} + \Delta G_k}{2}(t_{k+1} - t_k).$$

This quantity captures the magnitude of the glucose excursion above baseline and is expressed in units of mg/dL·min. The use of a fixed postprandial interval ensures comparability across repetitions. For each displacement $D_{i,f,r}$, the corresponding preceding excursion is taken as $\text{iAUC}_{i,f,r-1}$, enabling analysis of dependence between prior response magnitude and subsequent baseline shift.

**Statistical analysis.** Baseline displacement magnitudes were summarized using median and mean values and uncertainty in the median was estimated by bootstrap resampling. To assess whether observed displacements exceed local baseline variability, we compared the proportion of cases in which $|D|$ exceeds one or two pooled baseline standard deviations with the corresponding expectations under a Gaussian noise model. Because the dataset lacks paired reference glucose measurements, true sensor noise cannot be directly estimated. We therefore used a data-driven surrogate based on short-term pre-meal glucose increments, which yielded a robust variability scale of 3.41 mg/dL; baseline displacements were then evaluated relative to this empirical estimate. Associations between preceding postprandial excursions and subsequent baseline shifts were assessed using Pearson correlation coefficients, complemented by a within-subject permutation test to evaluate whether observed correlations could arise from random temporal ordering.

**Temporal dependence analysis.** Dependence between successive perturbations was assessed by computing Pearson correlation coefficients between displacement measures and preceding excursion magnitudes. For paired observations $(X_j, Y_j)$ corresponding to $\text{iAUC}_{i,f,r-1}$ and $|D_{i,f,r}|$, the correlation coefficient is defined as

$$r = \frac{\sum_j (X_j - \bar{X})(Y_j - \bar{Y})}{\sqrt{\sum_j (X_j - \bar{X})^2}\sqrt{\sum_j (Y_j - \bar{Y})^2}}.$$

Significance was assessed using the corresponding $t$-statistic

$$t = r\sqrt{\frac{N-2}{1-r^2}},$$

with $N-2$ degrees of freedom. Both signed displacements $D_{i,f,r}$ and absolute values $|D_{i,f,r}|$ were analyzed to distinguish directional effects from magnitude dependence.

To further characterize temporal dependence between successive perturbations, we implemented a one-lag regression and a surrogate-data test using only observed measurements. For each subject $i$, food condition $f$ and repetition $r$, the signed baseline displacement $D_{i,f,r}$ was modeled as a function of the preceding postprandial excursion and baseline level according to

$$D_{i,f,r} = \beta_0 + \beta_1 \, \text{iAUC}_{i,f,r-1} + \beta_2 \, G_{i,f,r-1}^* + \varepsilon_{i,f,r},$$

where $\text{iAUC}_{i,f,r-1}$ is the incremental area under the curve (mg/dL·min) in the previous repetition and $G_{i,f,r-1}^*$ is the preceding baseline (mg/dL). Parameters were estimated by ordinary least squares and statistical significance was assessed using the associated $t$-statistics and corresponding probability values derived from the Student distribution with $N-p$ degrees of freedom, where $N$ is the number of paired observations and $p$ the number of parameters.

In parallel, we performed a within-subject permutation test to evaluate whether the observed association between $\text{iAUC}_{i,f,r-1}$ and the absolute displacement $|D_{i,f,r}|$ could arise from random temporal ordering. For each subject, the sequence of iAUC values was randomly permuted while keeping the corresponding $|D|$ values fixed, thereby preserving marginal distributions and subject-level structure while disrupting temporal pairing. This procedure was repeated $B = 5000$ times to generate a null distribution of correlation coefficients $r^{(b)}$. The empirical two-sided probability value was computed as

$$p = \frac{1}{B+1}\sum_{b=1}^{B} \mathbf{1}\left(|r^{(b)}| \geq |r_{\text{obs}}|\right),$$

where $r_{\text{obs}}$ denotes the observed correlation. This approach provides a data-driven assessment of temporal dependence without assuming a specific noise model.

All computations were performed using Python with the libraries pandas, NumPy, SciPy and Matplotlib.

RESULTS

We report quantitative analysis of baseline glucose dynamics across repeated identical meal challenges in strictly normoglycemic individuals, focusing on the magnitude of baseline displacement, its relation to local variability and its dependence on preceding perturbations.

**Baseline variation**. Continuous-time glucose trajectories aligned to meal onset (Figure A) show variability in postprandial responses across repeated identical inputs, while pre-meal segments (Figure B) reveal that baseline levels differ across repetitions prior to perturbation. Across 67 repeated meal pairs, the median absolute baseline displacement was 8.09 mg/dL (mean 9.62 mg/dL; 95% bootstrap interval 4.81–10.90 mg/dL). Local baseline variability, estimated within the pre-meal window, had a median value of 3.33 mg/dL. When comparing displacement magnitude to local variability, 68.7% of pairs exceeded one pooled standard deviation and 50.7% exceeded two, both proportions significantly higher than Gaussian expectations (binomial test, $p < 0.001$ for both comparisons). Because paired reference measurements are not available, sensor noise was approximated using a data-driven surrogate based on short-term pre-meal increments, yielding a robust variability scale of 3.41 mg/dL; baseline displacements remained substantially larger than this empirical estimate. Sensitivity analyses using alternative pre-meal windows of 10–20 minutes yielded similar displacement magnitudes and preserved the association with prior excursions, indicating robustness to baseline definition. The distribution of |D| (Figure C) and subject-level variability (Figure D) indicate that non-zero baseline shifts are consistently observed across individuals. These results suggest that variability is expressed at the level of the baseline itself and not confined to short-scale fluctuations.

**History dependence**. The relationship between prior postprandial excursions and subsequent baseline displacement was assessed using incremental area under the curve (iAUC). The correlation between preceding excursion magnitude and absolute displacement was positive ($r = 0.354$, $p < 0.001$), indicating that larger perturbations are followed by larger baseline shifts. A within-subject permutation test confirmed that this association exceeds what is expected from random temporal ordering (Figure E), supporting the presence of structured temporal dependence. In contrast, the relationship between preceding excursion and signed displacement was weaker and not statistically significant, as shown by the dispersion of values in the lagged scatter plot (Figure F). Additional lagged regression analysis showed that the size of the preceding postprandial excursion is associated with the magnitude of the subsequent baseline displacement, such that larger excursions are followed by larger shifts in baseline levels. However, this relationship does not extend to the sign of the displacement: prior excursions do not predict whether the baseline will increase or decrease. Accordingly, successive baseline changes are not random in size, but their direction remains variable. This indicates that glycemic dynamics exhibit structured dependence in amplitude without a consistent directional response, meaning that prior perturbations influence how much the system changes, but not the direction in which it moves. Together, these findings indicate that successive perturbations influence subsequent system states in a structured manner, consistent with dependence across events rather than independent fluctuations.

Overall, our results show that baseline glucose levels vary across repeated identical conditions and that these variations exceed local variability estimates. The magnitude of baseline displacement is associated with prior excursions and is not attributable to random ordering. These findings indicate that glycemic dynamics involve structured adjustments across successive perturbations rather than independent fluctuations around a fixed baseline.

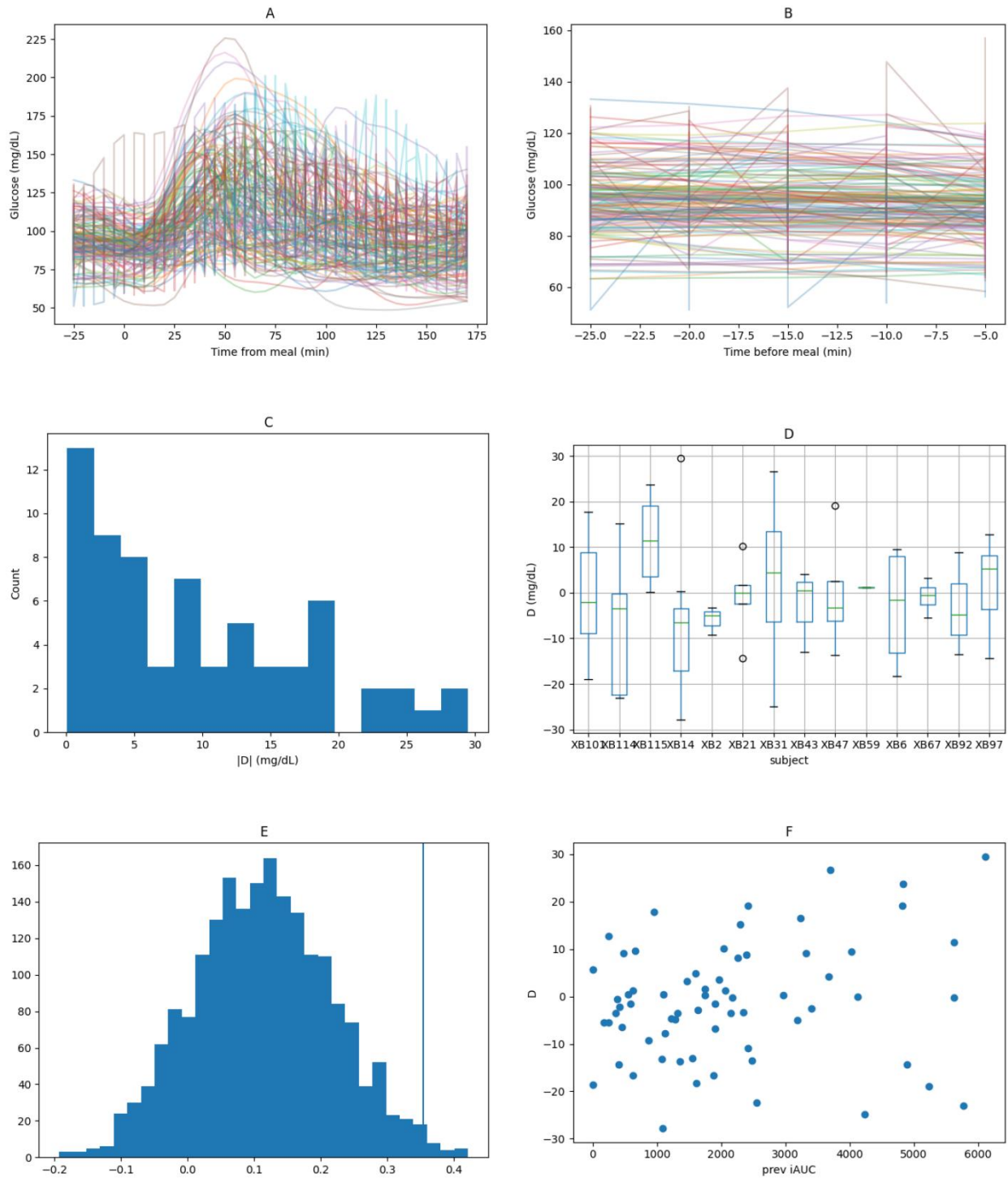

**Figure.** Non-random baseline displacement in repeated glycemic responses under controlled meal challenges.
(A) Glucose trajectories for repeated identical meal challenges plotted against continuous time relative to meal onset. Each curve represents one repetition within a subject and food condition. Despite identical inputs, trajectories differ in amplitude and recovery, indicating variability in postprandial responses.
(B) Pre-meal glucose segments in the interval −25 to −5 minutes. Differences across traces reveal that baseline levels vary across repetitions prior to perturbation.
(C) Distribution of absolute baseline displacement | $D$ |, defined as the absolute difference between consecutive pre-meal baselines within the same subject and food condition. The y-axis reports the number of repeated meal pairs. The breadth of the distribution indicates that baseline shifts frequently exceed the range of short-scale fluctuations observed within pre-meal intervals, reflecting variability at the level of the baseline rather than within-state noise.

(D) Subject-level distribution of baseline displacement $D$. The x-axis reports individual subjects. Distributions indicate that shifts are consistently centered away from zero across individuals, with variability in both dispersion and central tendency. The overall distribution differs from a zero-centered process, with a median displacement significantly different from zero (sign test, $p < 0.001$), indicating that baseline changes are not confined to symmetric noise around a fixed level.
(E) Null distribution of correlation coefficients obtained from within-subject permutation of the temporal pairing between previous postprandial excursion (incremental area under the curve, mg/dL·min) and subsequent displacement magnitude | $D$ | (mg/dL). The x-axis reports the correlation coefficient $r$ computed for each permuted dataset, while the y-axis shows the number of permutations. The observed correlation is shown relative to this distribution and lies in its upper tail, indicating that the association between prior excursion magnitude and subsequent baseline displacement exceeds what is expected from random temporal ordering.
(F) Relationship between previous excursion (mg/dL·min) on the x-axis and signed baseline displacement $D$ between consecutive baselines (mg/dL) on the y-axis. Points are broadly dispersed across positive and negative values of $D$ over the full range of excursions, indicating that, while larger perturbations are associated with larger subsequent changes in magnitude, the direction of the shift is not determined by the preceding excursion.

CONCLUSIONS

We asked whether glycemic regulation under repeated identical perturbations can be described as a return to a fixed baseline or whether the baseline itself varies across successive events. We compared pre-perturbation baseline estimates across repeated meal challenges within the same individuals, quantifying their differences and relating them to preceding postprandial excursions. We found that baseline displacements are consistently present, exceed both local baseline variability and an empirical short-term variability estimate derived from pre-meal glucose increments. The magnitude of these displacements is associated with the size of the previous excursion and this association persists under within-subject permutation testing, indicating that it cannot be explained by random temporal ordering. Still, lagged analyses show that prior perturbations do not determine the direction of the subsequent shift. These observations imply that baseline levels are not invariant but change across repetitions and these changes are structured in magnitude while not following a fixed signed rule. This suggests that glycemic regulation could proceed through successive adjustments rather than strict restoration of a single equilibrium.

Compared with existing techniques quantifying dispersion around a presumed stable baseline, we introduce a directly computable quantity, namely, the baseline displacement, which captures variation at the level of the equilibrium estimate. This quantity, derived from standard continuous glucose monitoring data without requiring additional measurements, can be evaluated independently of specific model assumptions. Conventional summaries do not distinguish between fluctuations within a state and shifts between states, whereas our analysis separates these components and allows direct assessment of whether repeated observations remain centered on a constant reference or evolve across repetitions. Further, the inclusion of surrogate and lagged analyses enables evaluation of temporal dependence without imposing a predefined noise model, providing a complementary dimension of characterization within existing analytical pipelines.

Our study has limitations.
Baseline estimation was limited to the interval [-25, -5] minutes due to data availability, differing from standard longer pre-meal windows and potentially introducing bias in the baseline definition. The displacement metric is mathematically well defined, but its interpretation assumes independence between consecutive baseline windows, whereas physiological processes may introduce overlap across adjacent periods. The comparison with Gaussian noise uses theoretical probabilities and modeling assumptions rather than empirically measured sensor noise. Correlation analysis identifies association between prior excursions and subsequent displacement magnitude but does not establish causality. Visualizations patterns rely on ordering of repetitions rather than continuous temporal trajectories. Still, our analysis was conducted on a single publicly available dataset, so all results rely on the Stanford CGM database alone. The cohort size is limited (n = 14), but the study is based on a repeated-measures design that increases inferential resolution by leveraging within-subject comparisons under controlled conditions. Each subject contributes multiple repeated identical meal challenges, yielding 67 paired observations of baseline displacement. This structure allows each individual to serve as their own control, reducing inter-individual variability and isolating the quantities of interest. The main measures are defined at the level of repeated events rather than subjects and are therefore estimated from a larger number of observations than the number of participants alone would suggest. The effect is observed across subjects, as indicated by subject-level distributions and is not driven by a small subset of individuals. The use of strict normoglycemic inclusion criteria further limits heterogeneity and ensures that the observed patterns are not attributable to overt metabolic dysregulation, providing sufficient resolution to evaluate deviations from a fixed-baseline description and their dependence on prior perturbations.

Testable experimental hypotheses follow from our approach. First, if baseline variability exceeds local noise, then in independent cohorts the proportion of cases where | $D_n$ |> $\sigma^{\text{baseline}}$ should remain greater than 0.317 and the proportion

where $|D_n| > 2\sigma^{\text{baseline}}$ should remain greater than 0.0455; these thresholds provide explicit statistical expectations under a Gaussian noise model.

Second, if prior excursions influence subsequent states, then the magnitude of displacement should scale with the preceding incremental area, such that $|D_n| = k\,\text{iAUC}_{n-1}^{\alpha}$ with $\alpha > 0$ and correlation coefficients between $|D_n|$ and $\text{iAUC}_{n-1}$ should remain positive and reproducible across datasets.

Third, if the process is not memoryless, then sequences should exhibit non-zero lag-1 autocorrelation, $\rho(1) = \text{corr}(D_n, D_{n+1}) > 0$, measurable over repeated perturbations.

Fourth, if baseline shifts reflect cumulative adaptation, then in continuous monitoring data the cumulative sum $S_N = \sum_{n=1}^{N} D_n$ should deviate from zero beyond random walk expectations over multi-day intervals.

Future research should evaluate these predictions in larger populations and in continuous free-living datasets with dense temporal sampling, allowing assessment of long-range temporal structure and validation of autocorrelation properties. Sensitivity analyses should test alternative baseline definitions and integration windows to determine robustness of displacement estimates. Comparative analyses across metabolic states may determine whether scaling exponents or displacement distributions differ systematically between normoglycemic and dysregulated conditions.

Our analysis enables refinement of glucose data interpretation by introducing a complementary descriptor derived from standard recordings. By quantifying successive baseline differences, it allows comparison of repeated exposures under controlled conditions at the level of individual trajectories, rather than relying solely on aggregate summaries. This can support individualized profiling of response consistency across identical inputs and facilitate identification of subtle deviations undetected by conventional indicators. Our method can be implemented using existing continuous glucose monitoring outputs without requiring additional sensors or experimental modifications and can be incorporated into routine data processing pipelines. It also provides a basis for evaluating the reproducibility of responses within subjects, providing a quantitative criterion to assess stability across repeated conditions. Integration with longitudinal datasets may enable tracking of temporal changes in response patterns, supporting detailed monitoring of individual glucose dynamics over time. Unanswered questions include whether similar baseline displacement patterns persist in continuous free-living conditions, how they evolve over longer time scales and whether directional components of displacement can be predicted from prior states.

In conclusion, we analyzed repeated meal challenges in normoglycemic individuals and identified consistent differences in pre-meal glucose levels across repetitions that exceeded short-scale variability and were related to prior responses. These findings indicate that glycemic dynamics cannot be interpreted as independent fluctuations around a fixed baseline, but instead reflect a history-dependent process in which successive perturbations influence subsequent system states. Accordingly, glucose time series are better described as evolving across repeated perturbations through measurable baseline displacements that depend on prior responses, rather than as fluctuations around a fixed reference level, providing a measurable signature of structured, non-random dynamics in glycemic regulation.

**DECLARATIONS**

**Ethics approval and consent to participate.** This research does not contain any studies with human participants or animals performed by the Author.
**Consent for publication.** The Author transfers all copyright ownership, in the event the work is published. The undersigned author warrants that the article is original, does not infringe on any copyright or other proprietary right of any third part, is not under consideration by another journal and has not been previously published.
**Availability of data and materials.** All data and materials generated or analyzed during this study are included in the manuscript. The Author had full access to all the data in the study and took responsibility for the integrity of the data and the accuracy of the data analysis.
**Disclaimer**. The views expressed are those of the author and do not necessarily reflect those of the affiliated institutions.
**Competing interests.** The Author does not have any known or potential conflict of interest including any financial, personal or other relationships with other people or organizations within three years of beginning the submitted work that could inappropriately influence or be perceived to influence their work.
**Funding.** This research did not receive any specific grant from funding agencies in the public, commercial or not-for-profit sectors.
**Acknowledgements:** none.
**Authors' contributions.** The Author performed: study concept and design, acquisition of data, analysis and interpretation of data, drafting of the manuscript, critical revision of the manuscript for important intellectual content, statistical analysis, obtained funding, administrative, technical and material support, study supervision.
**Declaration of generative AI and AI-assisted technologies in the writing process.** During the preparation of this work, the author used ChatGPT 5.3 to assist with data analysis and manuscript drafting and to improve spelling, grammar and general editing. After using this tool, the author reviewed and edited the content as needed, taking full responsibility for the content of the publication.

**REFERENCES**

1) Ajjan, R. A., T. Battelino, X. Cos, S. Del Prato, J. C. Philips, L. Meyer, J. Seufert, and S. Seidu. 2024. "Continuous Glucose Monitoring for the Routine Care of Type 2 Diabetes Mellitus." *Nature Reviews Endocrinology* 20 (7): 426–440. https://doi.org/10.1038/s41574-024-00973-1.
2) Chai, T. Y., S. Leathwick, M. M. Agarwal, D. B. Sacks, and D. Simmons. 2025. "Continuous Glucose Monitoring in Gestational Diabetes Mellitus: Hope or Hype?" *Diabetes Research and Clinical Practice* 227: 112389. https://doi.org/10.1016/j.diabres.2025.112389.
3) Ehrlich, R., A. Hendler-Neumark, V. Wulf, D. Amir, and G. Bisker. 2021. "Optical Nanosensors for Real-Time Feedback on Insulin Secretion by Beta-Cells." *Small* 17 (30): e2101660. https://doi.org/10.1002/smll.202101660.
4) Ferreira, R. O. M., T. Trevisan, E. Pasqualotto, M. P. Chavez, B. F. Marques, R. N. Lamounier, and S. van de Sande-Lee. 2024. "Continuous Glucose Monitoring Systems in Noninsulin-Treated People with Type 2 Diabetes: A Systematic Review and Meta-Analysis of Randomized Controlled Trials." *Diabetes Technology & Therapeutics* 26 (4): 252–262. https://doi.org/10.1089/dia.2023.0390.
5) Gregorio, K. C. R., C. P. Laurindo, and U. F. Machado. 2021. "Estrogen and Glycemic Homeostasis: The Fundamental Role of Nuclear Estrogen Receptors ESR1/ESR2 in Glucose Transporter GLUT4 Regulation." *Cells* 10 (1): 99. https://doi.org/10.3390/cells10010099.
6) Guo, H., L. Pan, Q. Wu, L. Wang, Z. Huang, J. Wang, L. Wang, X. Fang, S. Dong, Y. Zhu, and Z. Liao. 2025. "Type 2 Diabetes and the Multifaceted Gut-X Axes." *Nutrients* 17 (16): 2708. https://doi.org/10.3390/nu17162708.
7) Hall, H., D. Perelman, A. Breschi, P. Limcaoco, R. Kellogg, T. McLaughlin, and M. Snyder. 2018. "Glucotypes Reveal New Patterns of Glucose Dysregulation." *PLoS Biology* 16 (7): e2005143. https://doi.org/10.1371/journal.pbio.2005143.
8) Horgan, R., Y. Hage Diab, M. Fishel Bartal, B. M. Sibai, and G. Saade. 2024. "Continuous Glucose Monitoring in Pregnancy." *Obstetrics & Gynecology* 143 (2): 195–203. https://doi.org/10.1097/AOG.0000000000005374.
9) Jung, I. R., F. Anokye-Danso, S. Jin, R. S. Ahima, and S. F. Kim. 2022. "IPMK Modulates Hepatic Glucose Production and Insulin Signaling." *Journal of Cellular Physiology* 237 (8): 3421–3432. https://doi.org/10.1002/jcp.30827.
10) Lewis, G. F., A. C. Carpentier, S. Pereira, M. Hahn, and A. Giacca. 2021. "Direct and Indirect Control of Hepatic Glucose Production by Insulin." *Cell Metabolism* 33 (4): 709–720. https://doi.org/10.1016/j.cmet.2021.03.007.
11) Lu, J., Z. Ying, P. Wang, M. Fu, C. Han, and M. Zhang. 2024. "Effects of Continuous Glucose Monitoring on Glycaemic Control in Type 2 Diabetes: A Systematic Review and Network Meta-Analysis of Randomized Controlled Trials." *Diabetes, Obesity and Metabolism* 26 (1): 362–372. https://doi.org/10.1111/dom.15328.
12) Norton, L., C. Shannon, A. Gastaldelli, and R. A. DeFronzo. 2022. "Insulin: The Master Regulator of Glucose Metabolism." *Metabolism* 129: 155142. https://doi.org/10.1016/j.metabol.2022.155142.
13) Panigrahi, A., and S. Mohanty. 2023. "Efficacy and Safety of HIMABERB Berberine on Glycemic Control in Patients with Prediabetes: Double-Blind, Placebo-Controlled, and Randomized Pilot Trial." *BMC Endocrine Disorders* 23 (1): 190. https://doi.org/10.1186/s12902-023-01442-y.
14) Scoditti, E., S. Sabatini, F. Carli, and A. Gastaldelli. 2024. "Hepatic Glucose Metabolism in the Steatotic Liver." *Nature Reviews Gastroenterology & Hepatology* 21 (5): 319–334. https://doi.org/10.1038/s41575-023-00888-8.
15) Tao, Y., Q. Jiang, and Q. Wang. 2022. "Adipose Tissue Macrophages in Remote Modulation of Hepatic Glucose Production." *Frontiers in Immunology* 13: 998947. https://doi.org/10.3389/fimmu.2022.998947.
16) Wasserman, Edward A., Odysseus R. P. Orr, and Sophia Li. 2026. "Variability, Stability and the Law of Effect." *Journal of Experimental Psychology: Animal Learning and Cognition*. Published April 6, 2026.
17) Wu, Y., B. Ehlert, A. A. Metwally, et al. 2025. "Individual Variations in Glycemic Responses to Carbohydrates and Underlying Metabolic Physiology." *Nature Medicine* 31: 2232–2243. https://doi.org/10.1038/s41591-025-03719-2.
18) Xie, B., P. M. Nguyen, and O. Idevall-Hagren. 2019. "Feedback Regulation of Insulin Secretion by Extended Synaptotagmin-1." *FASEB Journal* 33 (4): 4716–4728. https://doi.org/10.1096/fj.201801878R.
19) Zahalka, S. J., R. J. Galindo, V. N. Shah, and C. C. Low Wang. 2024. "Continuous Glucose Monitoring for Prediabetes: What Are the Best Metrics?" *Journal of Diabetes Science and Technology* 18 (4): 835–846. https://doi.org/10.1177/19322968241242487.
20) Zhang, Y., L. Liu, C. Wei, X. Wang, R. Li, X. Xu, Y. Zhang, G. Geng, K. Dang, Z. Ming, X. Tao, H. Xu, X. Yan, J. Zhang, J. Hu, and Y. Li. 2023. "Vitamin K2 Supplementation Improves Impaired Glycemic Homeostasis and Insulin Sensitivity for Type 2 Diabetes through Gut Microbiome and Fecal Metabolites." *BMC Medicine* 21 (1): 174. https://doi.org/10.1186/s12916-023-02880-0.